\documentclass[journal=nalefd,manuscript=article]{achemso}
\usepackage{graphicx}
\usepackage{bm}
\usepackage{color}
\usepackage{amsmath,amsfonts,amssymb,wasysym}
\usepackage{cuted}
\usepackage{graphicx}
\usepackage{siunitx}
\usepackage[english]{babel}
\usepackage{color}
\usepackage{booktabs,longtable}
\usepackage{natbib}
\usepackage{hyperref}

\title{Orbital to charge current conversion in copper oxide heterostructures.}

\author{S. Vojkovic}
\affiliation{Facultad de Física, Pontificia Universidad Católica de Chile, Avenida Vicuña Mackenna 4860, Casilla 306, Santiago, Chile}
\alsoaffiliation{Centro de Nanociencia y Nanotecnología (CEDENNA), Avenida Manuel Rodriguez Sur 415, Santiago, Chile}

\author{K. Cancino}
\affiliation{Departamento de F\'\i sica, Universidad de Santiago de Chile, USACH, Av. Victor Jara 3493, Santiago, Chile}
\alsoaffiliation{Centro de Nanociencia y Nanotecnología (CEDENNA), Avenida Manuel Rodriguez Sur 415, Santiago, Chile}

\author{G. Rodríguez}
\affiliation{Departamento de F\'\i sica, Universidad de Santiago de Chile, USACH, Av. Victor Jara 3493, Santiago, Chile}
\alsoaffiliation{Centro de Nanociencia y Nanotecnología (CEDENNA), Avenida Manuel Rodriguez Sur 415, Santiago, Chile}

\author{E. Burgos}
\affiliation{Departamento de F\'\i sica, Universidad de Santiago de Chile, USACH, Av. Victor Jara 3493, Santiago, Chile}
\alsoaffiliation{Departamento de física, Facultad de Ciencias, Universidad de Chile, Las Palmeras 3425, Ñuñoa, Santiago, Chile}

\author{G. Herrera}
\affiliation{Departamento de F\'\i sica, Universidad de Santiago de Chile, USACH, Av. Victor Jara 3493, Santiago, Chile}

\author{C. Gonzalez-Fuentes}
\affiliation{Facultad de Física, Pontificia Universidad Católica de Chile, Avenida Vicuña Mackenna 4860, Casilla 306, Santiago, Chile}

\author{J. Palma}
\affiliation{Centro de Investigación en Ingeniería de Materiales CIIMAT, Universidad Central de Chile, Santa Isabel 1186, 8330601 Santiago, Chile}
\alsoaffiliation{Centro de Nanociencia y Nanotecnología (CEDENNA), Avenida Manuel Rodriguez Sur 415, Santiago, Chile}

\author{T. V. M. Sreekanth}
\affiliation{School of Mechanical Engineering, Yeungnam University, Gyeongsan-si 38541, Republic of Korea}

\author{J. Denardin}
\affiliation{Facultad de Física, Pontificia Universidad Católica de Chile, Avenida Vicuña Mackenna 4860, Casilla 306, Santiago, Chile}
\alsoaffiliation{Centro de Nanociencia y Nanotecnología (CEDENNA), Avenida Manuel Rodriguez Sur 415, Santiago, Chile}

\author{R. L. Rodríguez-Suárez}
\affiliation{Facultad de Física, Pontificia Universidad Católica de Chile, Avenida Vicuña Mackenna 4860, Casilla 306, Santiago, Chile}

\author{S. Oyarz\'un}

\affiliation{Departamento de F\'\i sica, Universidad de Santiago de Chile, USACH, Av. Victor Jara 3493, Santiago, Chile}
\alsoaffiliation{Centro de Nanociencia y Nanotecnología (CEDENNA), Avenida Manuel Rodriguez Sur 415, Santiago, Chile}
\email{simon.oyarzun@usach.cl}
\begin{document}
\maketitle

\begin{abstract}
We investigate the orbital-to-charge current conversion in Co$_{40}$Fe$_{40}$B$_{20}$\textbar CuO bilayers as a function of CuO thickness, employing orbital pumping via ferromagnetic resonance. The dynamic injection of orbital angular momentum into the CuO layer generates a transverse voltage through the Inverse Orbital Hall Effect (IOHE). By systematically varying the CuO thickness from 2 nm to 30 nm, we observe a pronounced dependence of the IOHE-induced voltage on the CuO layer thickness, indicating efficient orbital-to-charge conversion. These results highlight the key role of the orbital degree of freedom in orbitronics and provide insights into the potential of transition-metal oxides for next-generation orbitronic devices.
\end{abstract}

Spintronic devices rely on three key operations: the generation, manipulation, and detection of spin currents. One of the primary mechanisms to generate spin currents is from charge currents via the spin Hall effect (SHE) \cite{ikb,Hoffmann:2013el,Valenzuela:2006cs}. 
Spin currents can interact with local spins in ferromagnets, inducing magnetization dynamics, which is referred to as spin torque \cite{Slonczewski:1996wo,Berger:1996jd,Ralph:2008vb}. The reciprocal effect of spin torque is the spin pumping (SP), which refers to the emission of spin currents from precessing magnetization \cite{2005JAP97jC715A,2006ApPhL..88r2509S,Tserkovnyak:2002bm,Tserkovnyak:2002ju,Tserkovnyak:2005fr}. Both effects have been central to the development of spintronics. However, because SHE arises from spin-orbit coupling (SOC), efficient spintronics devices typically rely on heavy metals, such as Pt, W, and Ta. 
Although spin transport has received the most attention in spintronics, electrons in solids can carry both spin and angular momentum. Recent theoretical predictions and experimental studies  \cite{10.1103/physrevb.103.l020407,10.1103/physrevb.111.l140409,10.1038/s41928-024-01193-1,10.1038/s42005-021-00737-7,10.1063/5.0170654,10.1103/physrevb.109.014420} have demonstrated that a flow of orbital angular momentum (OAM) can be efficiently generated in light metals, such as Ti, Al, and oxidized Cu. This effect, known as the Orbital Hall effect (OHE), involves the generation of an orbital current in a normal metal (NM) perpendicular to the applied charge current. Orbital currents generated in metals in the absence of SOC can be absorbed by ferromagnets (FM) through the orbit-to-spin conversion mediated by SOC  \cite{10.1103/physrevb.103.l020407}. Thus, magnetization dynamics can also be driven by currents of OAM rather than spin currents \cite{2fl}. This so-called orbital torque can be viewed as an analogue of the spin torque, with a crucial distinction distinction that the latter requires SOC, since local magnetic moments in the FM are coupled directly to spin via exchange interaction. At the same time, OAM interacts with the magnetization indirectly through SOC. Onsager´s reciprocal relations guarantee the reciprocal effect of the orbital torque, in which an orbital current is pumped by magnetization precession. This phenomenon, denoted as orbital pumping (OP)  \cite{10.1103/physrevb.111.l140409}, has been reported recently in Ni/Ti bilayers by Hayashi et al.\cite{10.1038/s41928-024-01193-1}. Like spin pumping, orbital pumping involves the excitation of coherent magnetization precession by rf radiation in bilayers of ferromagnetic\textbar normal metal (FM\textbar NM) heterostructures. The precessing magnetization in the presence of SOC works as a pump, injecting both spin and orbital currents that in turn generate charge currents in the NM layer through the inverse spin Hall effect (ISHE) and inverse orbital Hall effect (IOHE), respectively.

In this work, we report a systematic investigation of the inverse orbital Hall effect (IOHE) in Co$_{40}$Fe$_{40}$B$_{20}$ (15~nm)$|$CuO($t_{\mathrm{CuO}}$) ferromagnet--normal metal (FM$|$NM) bilayers, with the CuO thickness varied from 2~nm to 30~nm. In FM/NM systems driven under ferromagnetic resonance (FMR), the precessing magnetization of the ferromagnet pumps orbital angular momentum into the normal metal, giving rise to a charge current via orbital-to-charge conversion mediated by the IOHE. By solving a drift--diffusion model that accounts for orbital angular momentum diffusion in the NM layer, we show that, under FMR conditions, the resulting charge current is dominated by the IOHE.

\begin{figure}[ht]
\centering
\includegraphics[width=\columnwidth]{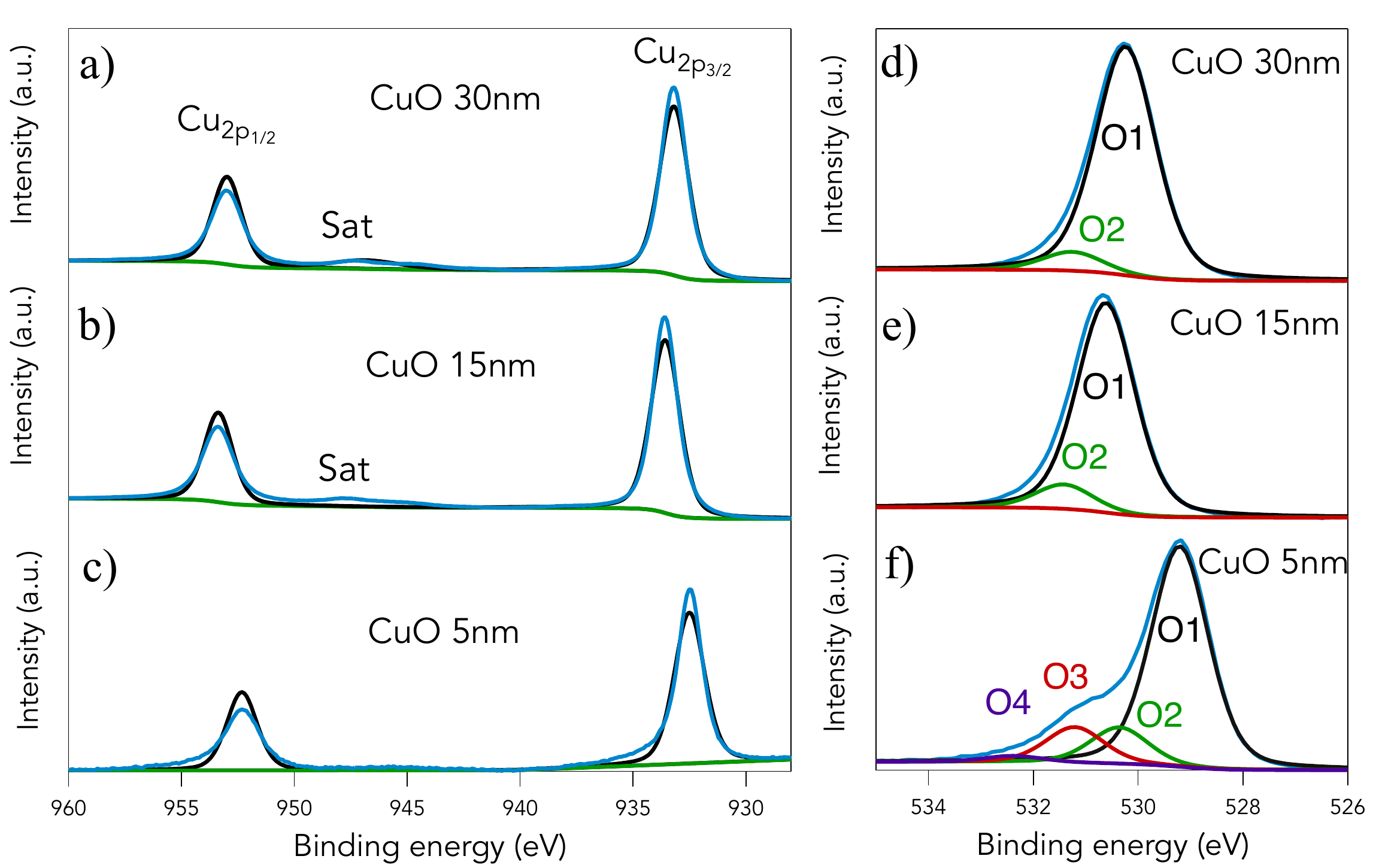}
\caption{\footnotesize High-resolution XPS spectra of CuO thin films with thicknesses of 5, 15, and 30~nm. (a–c) Cu 2$p$ core-level spectra for $t_{\mathrm{CuO}} = 5$, 15, and 30~nm, respectively, displaying the Cu 2$p_{3/2}$ and Cu 2$p_{1/2}$ components together with their satellite features. (d–f) Corresponding O 1$s$ core-level spectra for the same thicknesses, showing the deconvoluted oxygen components and their evolution with film thickness.}
\label{fig:XPS}
\end{figure}

The samples were grown by sputtering on mica substrate, where the base pressure of the system was $3 \times 10^{-7}$~mbar, and the pressure during the deposition was $3 \times 10^{-3}$~mbar by using a \SI{20}{sccm} of Ar flow. We used CoFeB and CuO sputtering targets from ACI alloys with a deposition rate of 2 nm/min and 0.75 nm/min,  respectively. The thickness of the samples was measured by AFM obtaining a height profile where it is possible to observe two steps corresponding to CoFeB and CuO layers. The rms roughness determined from the AFM images corresponds to around 1 nm. 
The elemental compositions of the samples were quantitatively compared via XPS (K-$\alpha$, Thermo Scientific, USA) using Al K$\alpha$ radiation (1486.6 eV). XPS analysis was performed on 5, 15, and 30 nm CuO thickness samples. The deconvoluted Cu-2$p$ spectra for these films are shown in Fig. \ref{fig:XPS} (a-c). The binding energy peaks at approximately 932.5/933.5/933.1 eV for the 5, 15, and 30 nm CuO thickness, respectively, corresponding to Cu $2p_{\mathrm {3/2}}$ are attributed to metallic copper and Cu-O bonds in CuO \cite{10.1038/s41467-020-16554-5,10.1039/d0nr02208j}. The Cu $2p_{\mathrm {1/2}}$ peaks appear at around 952.3/953.5/953 eV for the 5, 15, and 30 nm films, respectively, related to Cu $2p_{\mathrm {1/2}}$, indicating Cu$^{+}$ and Cu$^{2+}$ states \cite{10.1039/d0nr02208j,10.1016/j.est.2024.112021}. The O-1$s$ core level spectrum deconvolution was performed using Avantage software at the first deposition (5 nm, see Fig. \ref{fig:XPS} (f)), revealing four peaks at 529.2 eV (O1: M-O bond), 530.4 eV (O2: Cu-O), 531.2 eV (O3: surface hydroxyls), and 532.4 eV (O4: surface-adsorbed H$_{2}$O molecules containing O$_{2}$ species) \cite{10.1038/s41467-020-16554-5,10.1039/d0nr02208j,10.1016/j.est.2024.112021}. As the film thickness increased from 5 nm to 15 nm (Fig. \ref{fig:XPS} (e)) and then to 30 nm (Fig. \ref{fig:XPS} (d)), the initial peak (529.2 eV) and the final peak (532.4 eV) disappeared, while the other peaks shifted slightly to higher binding energies.

\begin{figure*}[ht]
\centering
\includegraphics[scale=0.35]{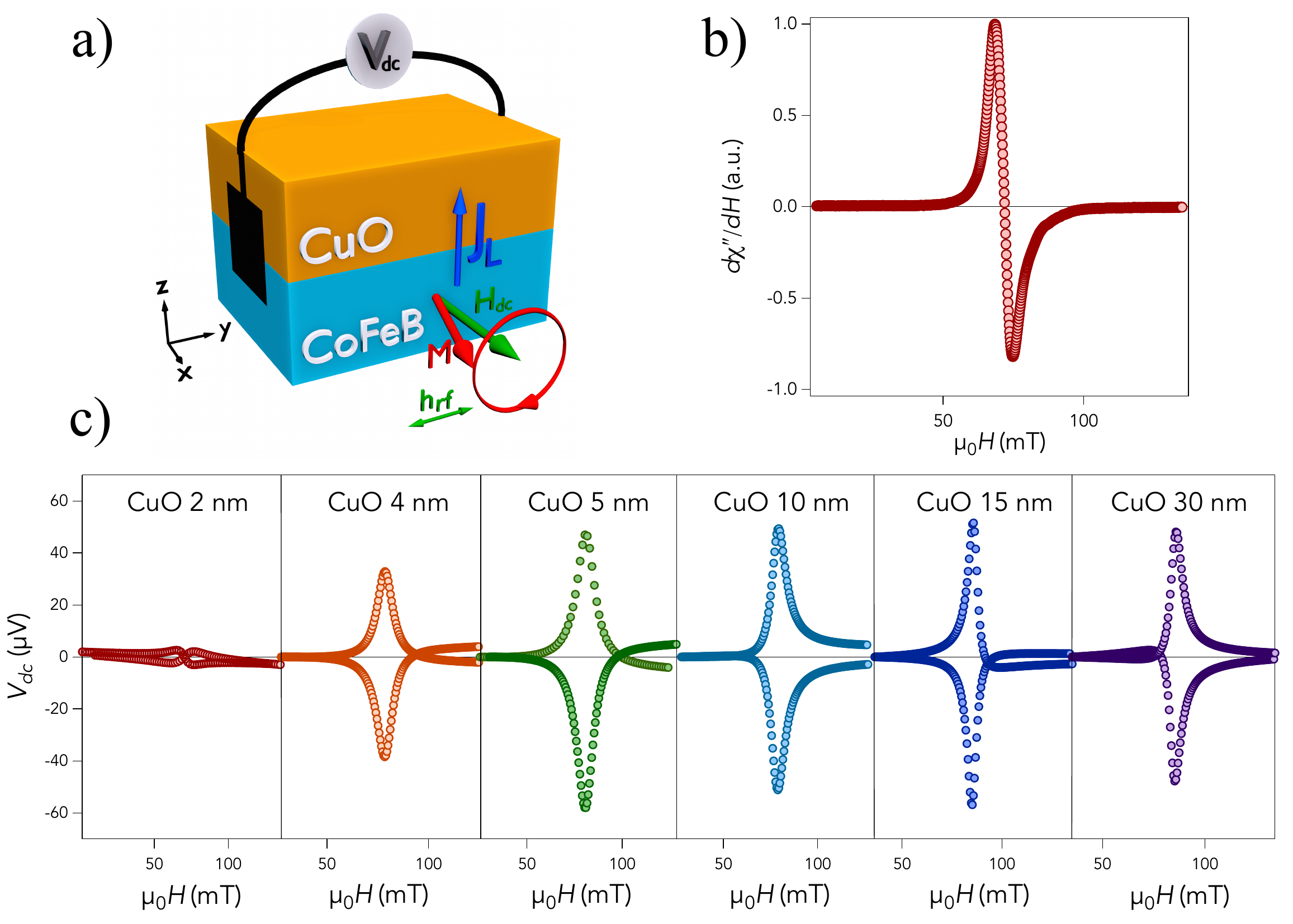}
\caption{\footnotesize(a) Schematic illustration of the CoFeB$|$CuO bilayer used for orbital pumping and $V_{\mathrm{IOHE}}$ measurements. A microwave field excites ferromagnetic resonance (FMR) in the CoFeB layer, leading to the injection of orbital angular momentum into the adjacent CuO layer. The resulting orbital-to-charge conversion generates a transverse voltage $V_{\mathrm{dc}}$, measured  across the sample. (b) Derivative of the FMR absorption signal as a function of the applied magnetic field, measured at a frequency of 9.8 GHz and a microwave power of 197 mW. (c) Measured inverse orbital Hall voltage \( V_{\mathrm{dc}} \) for the complete set of CoFeB$|$CuO samples with CuO thicknesses ranging from 2 to 30 nm.}
\label{fig:orbital pumping}
\end{figure*}

A natural route for the investigation of the orbital-to-charge current conversion is to measure the IOHE signal, similar to the ISHE detection in strong SOC materials such as Pt, W, Ta, and Pd \cite{Sinova:2015ic}. Here, we detect IOHE in CuO by using orbital-pumping (OP) technique in CoFeB(15 nm)/CuO ($t_{\mathrm{CuO}}$=2, 4, 5, 10, and 30 nm)  bilayers. For the measurements, the sample was placed at the center of a cylindrical Bruker X-band cavity, at a frequency $f$= 9.8 GHz, where the rf magnetic field is maximum and the rf electric field is minimum, to suppress galvanic effects induced by the electric field. In this configuration, the static magnetic field $H_{\rm{dc}}$ and the rf field lie in the plane of the film and are oriented perpendicular to each other, while the sample is rotated to record both the FMR spectra and the dc voltage induced by orbital-to-charge current conversion.

Figure~2(a) shows a schematic illustration of the CoFeB\textbar CuO bilayer used in the orbital pumping experiment to detect the inverse orbital Hall effect (IOHE). Under FMR conditions, the precessing magnetization injects an orbital current into the CuO layer, leading to orbital accumulation. A fraction of this orbital current $J_{L}$ is converted into a transverse charge current $J_{c}$ via the IOHE, given by
$\mathbf{J}_{c} = \frac{2e}{\hbar}\,\theta_{\mathrm{OH}}\,(\mathbf{J}_{L} \times \hat{\boldsymbol{\sigma}}_{L})$
where $\hat{\boldsymbol{\sigma}}_{L}$ is the orbital polarization and $\theta_{\mathrm{OH}}$ is the orbital Hall angle, which characterizes the efficiency of orbital-to-charge current conversion. As a consequence of the charge current induced by the IOHE, a dc voltage $V_{\mathrm{dc}}$ develops along the length of the sample and is recorded directly using a nanovoltmeter connected via copper wires to the electrodes shown in Fig.~2(a).

Figure~2(b) displays the derivative of the FMR spectrum for the CoFeB (15~nm)\textbar CuO (2~nm) sample, measured at a microwave power of 197~mW, while Fig.~2(c) presents the field-scan dc voltage $V_{\mathrm{dc}}$ for the complete set of samples. For the CuO (2~nm) sample, a fully asymmetric voltage signal with a small amplitude is observed. As the CuO thickness increases beyond 4~nm, the signal amplitude increases and a pronounced symmetric component develops, remaining nearly constant up to a thickness of 30~nm. The $V_{\mathrm{dc}}$ signal reverses sign upon a $180^{\circ}$ rotation of the sample (orbital polarization) with respect to the applied dc magnetic field $H_{\mathrm{dc}}$, and exhibits a linear dependence on the applied microwave power.

The raw $V_{\mathrm{dc}}$ data can be described as a combination of symmetric ($V_{\mathrm{Sym}}$) and antisymmetric ($V_{\mathrm{Asym}}$) components:
\begin{equation}
V_{\mathrm{dc}} =
 V_{\mathrm{Sym}}\,\frac{\Delta H^2}{(H - H_{R})^2+\Delta H^2} 
+ V_{\mathrm{Asym}}\,\frac{\Delta H(H - H_{R})}{(H - H_{R})^2 + \Delta H^2}
\label{eq:IOHE_fit}
\end{equation}
where $H_{R}$ is the resonance field, and $\Delta H$ is the half-width at half-maximum (HWHM) of the resonance peak. In FM metals, the $V_{\mathrm{Sym}}$ and $V_{\mathrm{Asym}}$ components originate from the anisotropic magnetoresistance (AMR) effect~\cite{2011PhRvB..83n4402A,2013PhRvB..88f4403R,Harder:2011ho}, whereas in FM\textbar NM bilayers the voltage generated by orbital pumping (OP) exhibits a purely symmetric contribution. Figure~4 shows the amplitude of the symmetric component $V_{\mathrm{Sym}}$ as a function of the CuO thickness $t_{\rm CuO}$. Except for the $t_{\rm CuO}=2$ nm sample, which displays a small asymmetric contribution, all other samples satisfy $V_{\mathrm{Asym}} \ll V_{\mathrm{Sym}}$. We attribute the dominant symmetric voltage signal to the IOHE in the CuO layer.

To describe the diffusion of the orbital current, we consider a model with a single orbital channel, that is, we disregard any orbital-to-spin and spin-to-orbital interconversion mediated by SOC inside the CuO layer~\cite{10.1103/physrevresearch.4.033037,10.1103/physrevb.109.014420}. In this framework, the orbital accumulation, represented by its chemical potential $\mu_L$, is generated solely by the OP process. The diffusion of orbital angular momentum is governed by
\begin{equation}
\frac{d^2 \mu_{L}}{dz^2} = \frac{\mu_{L}}{\lambda_{L}^2}
\end{equation}
where $\lambda_L$ is the orbital diffusion length \cite{10.1103/physrevb.109.014420}. 

The boundary conditions are determined by the continuity of the orbital current $J_L$ at the FM\textbar NM interface and the vanishing of the current at the outer boundary $z = d$, with $\sigma_{\mathrm{NM}}$ the longitudinal electrical conductivity of the NM layer:
\begin{subequations}
\begin{align}
J_{L}(0) = -\frac{\hbar\, \sigma_{\mathrm{NM}}}{2 e^2} \left.\frac{\partial \mu_{L}}{\partial z}\right|_{z=0}\label{eq3a}\\[12pt]
\left.\frac{\partial \mu_{L}}{\partial z}\right|_{z=d} = 0 \label{eq3b}
\end{align}
\end{subequations}

Note that the current in Eq.(\ref {eq3a}) is  expressed in units of angular momentum per (area$\cdot$time). 

The solution of Eqs.~(2) and (3) with the corresponding boundary conditions is
\begin{equation}
\mu_{L}(z) = 
\frac{\cosh{\!\left[\frac{(z - t_{\mathrm{CuO}})}{\lambda_{L}}\right]}}{\sinh{(t_{\mathrm{CuO}} / \lambda_{L})}}
\frac{2 e^2 \lambda_{L}}{\sigma_{\mathrm{NM}} \hbar} J_{L}(0)
\end{equation}

The orbital accumulation in NM drives an orbital current back into the FM through the interface,
\begin{equation}
J_{L}^{\mathrm{back}} = \frac{g_{L}}{4\pi} \,\mu_{L}(0)
\end{equation}
where $g_L$ is the orbital mixing conductance~\cite{10.48550/arxiv.2412.08340}. The total orbital current density at the interface is then
\begin{equation}
J_{L}(0) = J_{L}^{\mathrm{pump}} - J_{L}^{\mathrm{back}}
\end{equation}
with the $x$-polarized orbital pumped current density in the $z$ direction is
\begin{equation}
J_{L}^{\mathrm{pump}} = \frac{\hbar g_{L}}{4\pi M^2}\left(\mathbf{M} \times \frac{d\mathbf{M}}{dt}\right)_z
\end{equation}
Combining the above expressions, we obtain
\begin{equation}
J_{L}(0) = \left(1 + \beta_{L} g_{L}\right)^{-1} J_{L}^{\mathrm{pump}}
\end{equation}
where
\begin{equation}
\beta_{L} = \frac{\lambda_{L} e^2}{\sigma_{\mathrm{NM}} h} 
\frac{1}{\tanh{(t_{\mathrm{CuO}} / \lambda_{L})}}
\end{equation}

Under FMR, the precessing magnetization generates an orbital density current at the FM/NM interface ($z = 0$) given by

\begin{equation}
J_{L}(0) = 
\frac{\hbar \omega g_{L}^{\mathrm{eff}}}{4\pi}
\left(\frac{h_{\mathrm{rf}}}{\Delta H}\right)^2
\frac{(H_R + 4\pi M_S)}{(2H_R + 4\pi M_S)^2}
\frac{\omega}{\gamma}
\left[
\frac{\Delta H^2}{(H - H_R)^2 + \Delta H^2}
\right]
\end{equation}
where $g_{L}^{\mathrm{eff}} = g_{L}\bigl(1 + \beta_{L} g_{L}\bigr)^{-1}$ is the effective orbital  conductance at the interface that takes into account the pumped and blackflow, $\omega$ and $h_{\mathrm{rf}}$ are the frequency and amplitude of the driving microwave magnetic field, $M_S$ the saturation magnetization, $\Delta H$ is the HWHM and $\gamma$ the gyromagnetic ratio.

The orbital current injected at the interface diffuses inside the CuO layer with a characteristic diffusion length $\lambda_L$. Due to the inverse orbital Hall effect (IOHE), this diffusing orbital current generates a transverse charge current density,
\begin{equation}
J_{C}(z) = \theta_{\mathrm{OH}} \left(\frac{2e}{\hbar}\right) J_{L}(0)\,
\frac{\sinh\!\left(\frac{t_{\mathrm{CuO}} - z}{\lambda_L}\right)}{\sinh\!\left(\frac{t_{\mathrm{CuO}}}{\lambda_L}\right)}
\end{equation}
where $\theta_{\mathrm{OH}}$ is the orbital Hall angle. Integrating the charge current across the thickness gives the IOHE voltage
\begin{equation}
V_{\mathrm{IOHE}} =
\left(\frac{2e}{\hbar}\right)
\left(\frac{l \lambda_L}{\sigma_{\mathrm{NM}} t_{\mathrm{CuO}}}\right)\,
\theta_{\mathrm{OH}}\,
\tanh\!\left(\frac{t_{\mathrm{CuO}}}{2\lambda_L}\right)
J_{L}(0)
\label{eq:Viohe_final}
\end{equation}
where $l$ is the sample length and $\sigma_{\mathrm{NM}}$ is the longitudinal conductivity of CuO.

In order to determine the magnitude of $J_{L}(0)$ and the corresponding $V_{\mathrm{IOHE}}$, we performed broadband FMR measurements to obtain the Gilbert damping parameter $\alpha$ for each sample. A custom-built setup was employed, in which the sample was placed on a waveguide that transmits the microwave excitation to the CoFeB layer, while an external magnetic field was applied in the plane of the structure. The FMR signal, corresponding to the derivative of the magnetic susceptibility with respect to the external field, was recorded for microwave frequencies ranging from 6 to 16 GHz. From each spectrum, the peak-to-peak linewidth $\Delta H_{\mathrm{pp}}$ was extracted and plotted as a function of frequency. The data were fitted using the expression
\begin{equation}
\Delta H_{\mathrm{pp}} = \Delta H_0 + \frac{2}{\sqrt{3}}\left(\frac{\omega}{\gamma}\right)\alpha
\label{Gilbert}
\end{equation}
where $\Delta H_0$ accounts for inhomogeneities in the ferromagnetic layer. Figure~3(a) shows $\Delta H_{\mathrm{pp}}$ as a function of frequency for the complete set of samples. By fitting the data, we extracted the damping parameter $\alpha$, which is plotted as a function of CuO thickness in Fig.~3(b). 
\begin{figure}[ht]
\includegraphics[scale=0.27]{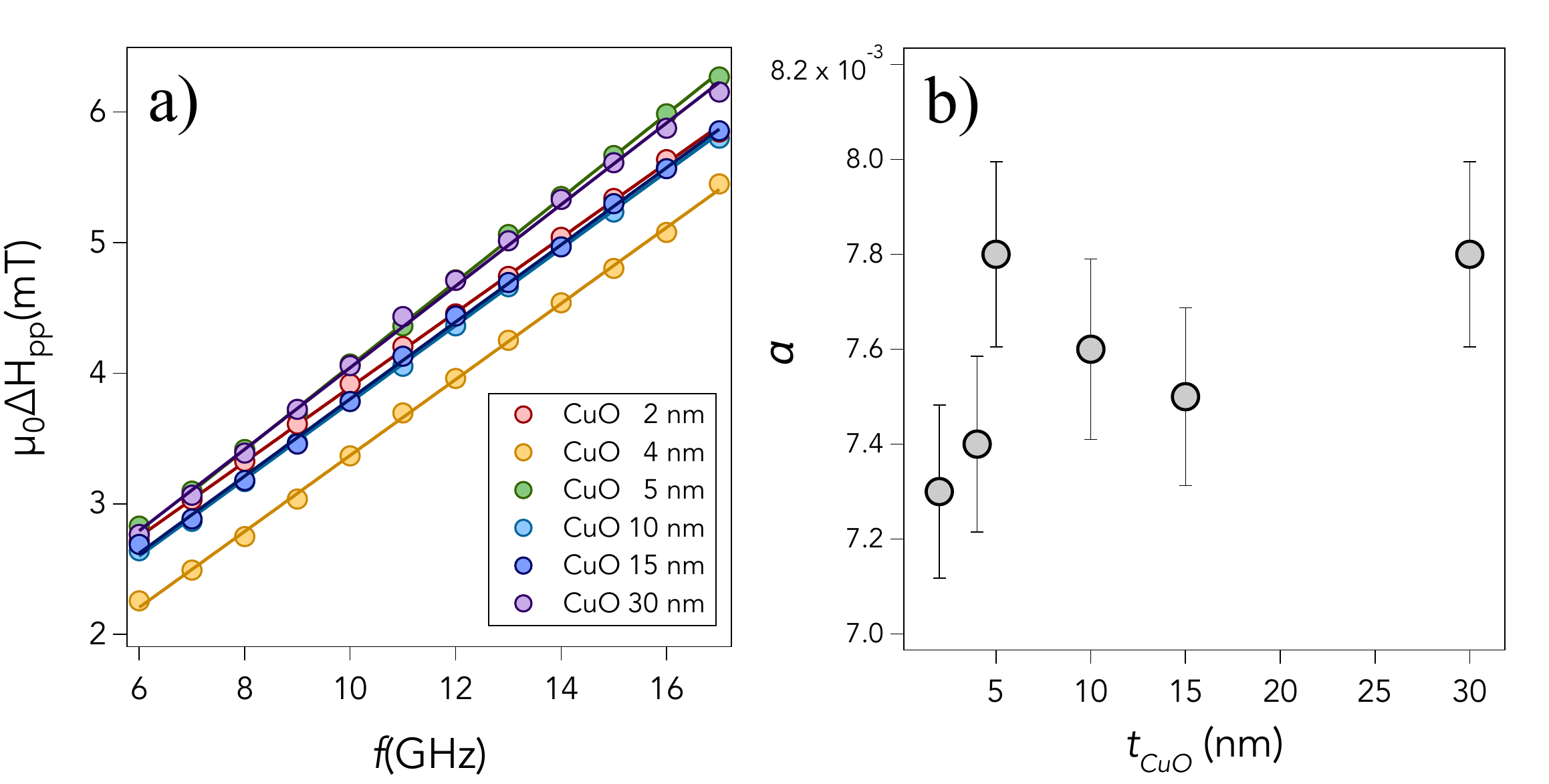}
\caption{\footnotesize (a) Peak-to-peak linewidth $\Delta H_{\mathrm{pp}}$ as a function of frequency for the complete set of samples. The solid lines represent fits using Eq.~(\ref{Gilbert}). (b) Extracted damping parameter $\alpha$ as a function of CuO thickness.}
\label{fig:damping}
\end{figure}
A small but finite increase of $\alpha$ with increasing CuO thickness is observed. This trend suggests that the CuO layer acts as an orbital angular momentum sink: orbital angular momentum pumped across the interface is dissipated in the CuO layer, providing an additional relaxation channel that slightly enhances the effective Gilbert damping \cite{10.1103/physrevb.111.l140409,10.1063/5.0292745}.

%

A key test of the orbital-pumping voltage theory $V_{\mathrm{IOHE}}$ is provided by its comparison with experimental data obtained by varying the thickness of the NM layer. From the damping parameter, we extract the effective orbital mixing conductance $g_{L}^{\mathrm{eff}}$ according to
\begin{equation}
\alpha = \gamma \frac{\hbar g_{L}^{\mathrm{eff}}}{4\pi M t_{\mathrm{FM}}}.
\end{equation}

\begin{figure}[htbp]
\centering
\includegraphics[scale=0.35]{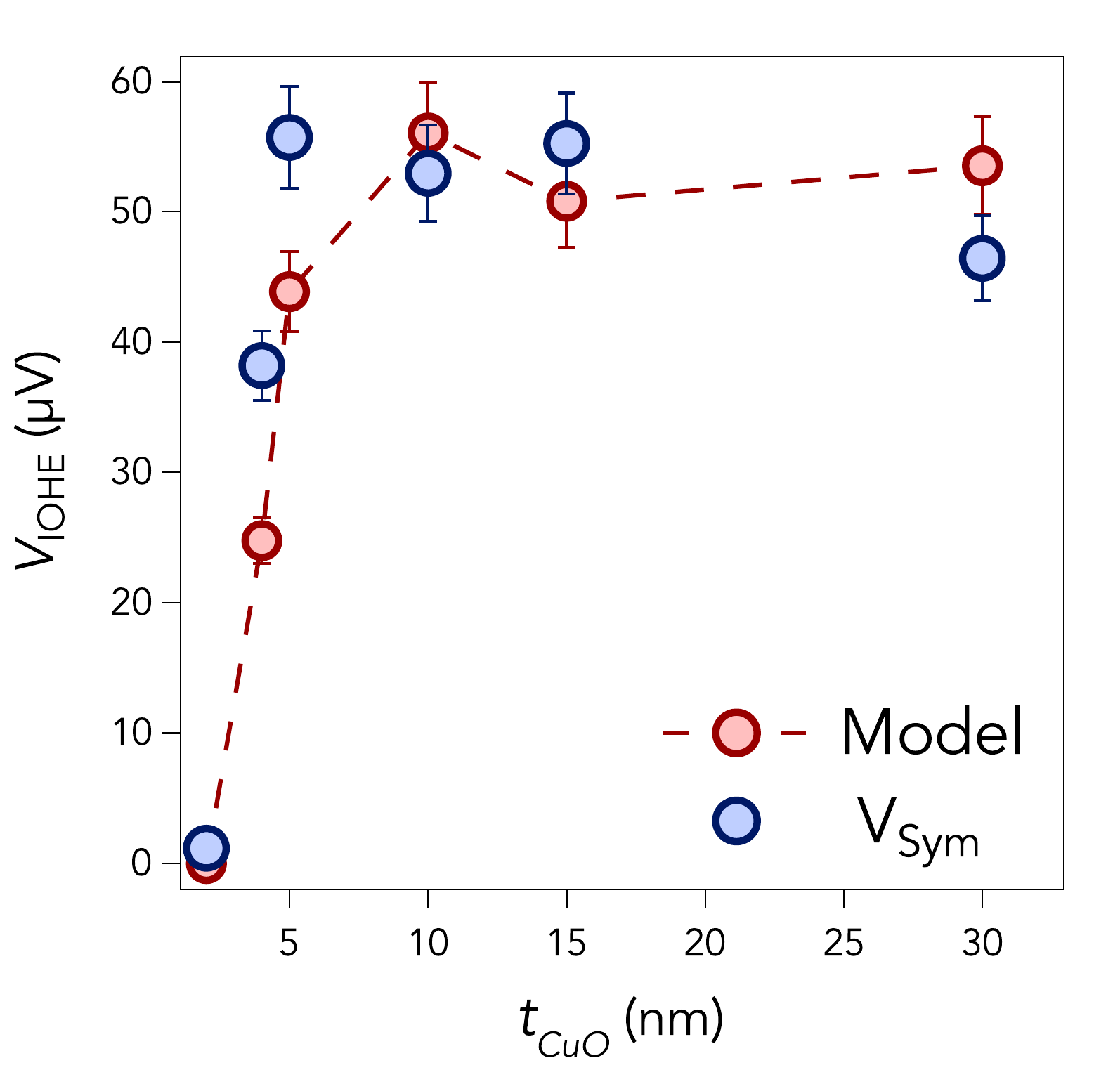}
\caption{\footnotesize Blue circles correspond to the symmetric voltage component $V_{\mathrm{Sym}}$ as a function of the CuO thickness $t_{\mathrm{CuO}}$, showing an increasing trend from 2 to 30~nm. Red circles correspond to the values obtained for the voltage using the orbital diffusion model.}
\label{fig:Vsym}
\end{figure}

The symmetric component ($V_{\mathrm{Sym}}$) of the $V_{\mathrm{dc}}$ data shown by the symbols in Fig.~\ref{fig:Vsym} is interpreted using Eq.~\ref{eq:Viohe_final}, together with the parameters listed in Table~I and those extracted from FMR measurements, namely $h_{\mathrm{rf}} = 1\,\mathrm{Oe}$ and $l = 4\,\mathrm{mm}$, corresponding to the sample length. The electrical conductivity $\sigma$ was measured for each sample using a four-point probe configuration based on the van der Pauw method, and to evaluate $V_{\mathrm{IOHE}}$ we assume $\sigma_{\mathrm{NM}} \simeq \sigma$. From this analysis, we obtain for the full set of samples an orbital Hall angle $\theta_{\mathrm{OH}} = 2\%$ and an orbital diffusion length $\lambda_L = 6\,\mathrm{nm}$.

The dc voltage detected in CoFeB\textbar CuO bilayers under ferromagnetic resonance exhibits a pronounced dependence on the CuO thickness and is dominated by a symmetric component. This behavior is consistent with a voltage generation mechanism governed by the diffusive propagation of nonequilibrium orbital angular momentum inside the CuO layer and its conversion into a transverse charge current via the inverse orbital Hall effect, as proposed in recent theoretical and experimental studies on orbital transport \cite{10.1103/physrevb.103.l020407,10.1103/physrevb.111.l140409}.

The thickness dependence of $V_{\mathrm{Sym}}$ is characterized by a monotonic increase for thin CuO layers, followed by saturation as the thickness exceeds the orbital diffusion length. Such behavior naturally emerges from the solution of the orbital diffusion equation with appropriate boundary conditions at the ferromagnet/normal-metal interface \cite{10.1103/physrevb.109.014420,10.1103/physrevresearch.4.033037}. The quantitative agreement between experiment and model over the entire thickness range indicates that orbital angular momentum transport in CuO is well described within a diffusive transport framework.

The extracted orbital diffusion length $\lambda_L = 6\,\mathrm{nm}$ is remarkably large for an oxide material and is comparable to values reported for light metals such as Ti and Al, where efficient orbital transport has been demonstrated despite weak spin--orbit coupling \cite{10.1038/s41928-024-01193-1,10.1063/5.0170654}. These results therefore extend the concept of robust diffusive orbital transport to fully oxidized transition-metal oxides.

The orbital Hall angle $\theta_{\mathrm{OH}} \approx 2\%$ further indicates that CuO acts as an efficient orbital-to-charge current converter. This magnitude is consistent with theoretical predictions and experimental observations of sizable orbital Hall responses in materials with moderate spin--orbit coupling \cite{10.1103/physrevb.103.l020407,10.1038/s42005-021-00737-7}. Importantly, the small but finite variation of the Gilbert damping parameter with CuO thickness, attributed to CuO acting as an orbital angular momentum sink, indicates that angular momentum is transferred across the interface without substantial spin absorption. This reinforces the interpretation that the detected voltage arises predominantly from orbital pumping and its conversion through the IOHE \cite{10.1103/physrevb.111.l140409}.

Taken together, these results establish CuO as a robust medium for diffusive orbital angular momentum transport. The combination of a finite orbital diffusion length and an appreciable orbital Hall angle positions CuO as an effective orbital channel in ferromagnet/oxide heterostructures, enabling angular momentum transfer and detection mediated primarily by orbital degrees of freedom without relying on heavy metals with strong spin--orbit coupling \cite{10.1103/physrevb.109.014420,10.48550/arxiv.2412.08340}.


\begin{table}[htbp]
\caption{Measured Gilbert damping parameter $\alpha$, electrical conductivity $\sigma$, effective orbital conductance $g_{L}^{\mathrm{eff}}$, and converted charge current $I_{\mathrm{c}}$ for the complete set of samples.}
\label{tab:1}
\centering
\begin{tabular}{ccccc}
\toprule
CuO thickness (nm) 
& $\alpha$ ($\times 10^{-3}$) 
& $\sigma \times 10^{5}$ ($\Omega\,\mathrm{m})^{-1}$ 
& $g_{L}^{\mathrm{eff}}$ ($\times 10^{18}\,\mathrm{m^{-2}}$) 
& $I_{\mathrm{c}}$ (nA) \\
\midrule
2  & 7.3 & 292 & 0.0 & 0.07 \\
4  & 7.4 & 370 & 1.1 & 33.6 \\
5  & 7.8 & 327 & 5.2 & 45.6 \\
10 & 7.6 & 265 & 3.1 & 43.9 \\
15 & 7.5 & 224 & 2.1 & 46.5 \\
30 & 7.8 & 152 & 5.2 & 39.9 \\
\bottomrule
\end{tabular}
\end{table}

The considerable value of the orbital diffusion length $\lambda_L$ reveals the robustness of orbital transport in CuO. Combined with the sizable orbital Hall angle, these results show that CuO can function as an efficient orbital transporter or “orbital filter” in FM\textbar NM heterostructures and in more complex multilayer stacks.

In summary, we demonstrate that the dc voltage generated in CoFeB\textbar CuO bilayers under ferromagnetic resonance is predominantly governed by orbital pumping and its conversion into a transverse charge current through the inverse orbital Hall effect. The thickness dependence of the voltage is quantitatively described by an orbital diffusion model, enabling the extraction of key transport parameters, including the orbital diffusion length and the orbital Hall angle. These results establish CuO as an efficient medium for diffusive orbital angular momentum transport and orbital-to-charge conversion in oxide-based heterostructures.

\section{Acknowledgments}
This work was funded by ANID CEDENNA CIA 250002, ANID Fondecyt 1241257, POSTDOC\_\,DICYT 042431CD\_\,Postdoc VRIIC, Fondecyt 1201491 and FONDEQUIP projects EQM180103  and EQM210088. S.V aknowledges support from ANID BECAS/Doctorado Nacional N° 21202132

\bibliography{referencesCuO.bib}
\end{document}